\input phyzzx
\def\e{\adveq\eqno{\rm (\chapterlabel\the\equanumber)}}

\def\adveq{\global\advance\equanumber by 1}
\def\myeq{{\rm \chapterlabel\the\equanumber}}
\def\rarrow{\rightarrow}

\def\twoline#1#2{\displaylines{\qquad#1\hfill(\adveq\myeq)\cr\hfill#2
\qquad\cr}}

\def\semidirect{\mathrel{\raise0.04cm\hbox{${\scriptscriptstyle |\!}$
\hskip-0.175cm}\times}}

\def\mod{\mathop{\rm mod}\nolimits}

\def\ref#1{$^{[#1]}$}

\def\r#1{$[\rm#1]$}

\def\e{\adveq\eqno{\rm (\chapterlabel\the\equanumber)}}

\def\adveq{\global\advance\equanumber by 1}
\def\myeq{{\rm \chapterlabel\the\equanumber}}
\def\rarrow{\rightarrow}

\def\twoline#1#2{\displaylines{\qquad#1\hfill(\adveq\myeq)\cr\hfill#2
\qquad\cr}}

\def\semidirect{\mathrel{\raise0.04cm\hbox{${\scriptscriptstyle |\!}$
\hskip-0.175cm}\times}}

\def\mod{\mathop{\rm mod}\nolimits}

\def\ref#1{$^{[#1]}$}

\def\r#1{$[\rm#1]$}

\def\mod {\, \rm mod\,}
\date{December, 2012}
\date{December, 2012}
\titlepage
\title{Generalized Rogers Ramanujan Identities from AGT Correspondence}
\author{Alexander A. Belavin$^{1,2,3}$ and Doron R. Gepner$^4$}
\vskip20pt
\line{$^1$ \it Landau Instiute for Theoretical Physics, Chernogolovka, Russia\hfill}
\line{$^2$ \it Institute for Information Transmission Problems, Moscow, Russia\hfill}
\line{$^3$ \it Moscow Institute of Physics and Technology, Dolgoprudny, Russia\hfill}
\line{$^4$ \it Department of Particle Physics, Weizmann Institute, Rehovot, Israel\hfill} 

\abstract
AGT correspondence and its generalizations attracted a great deal of attention recently.
In particular it was suggested that $U(r)$ instantons on $R^4/Z_p$ describe the conformal blocks
of the coset  ${\cal A}(r,p)=U(1)\times sl(p)_r\times {sl(r)_p\times sl(r)_n\over sl(r)_{n+p}}$,
where $n$ is a parameter.
Our purpose here is to describe Generalized Rogers Ramanujan (GRR) identities for these cosets,
which expresses the characters as certain $q$ series. 
We propose that such identities exist for the coset ${\cal A}(r,p)$ for all positive integers $n$
and all $r$ and $p$. 
We treat here the case of $n=1$ and $r=2$, finding GRR identities for all the characters. 
\endpage

Our interest here is to describe some generalized Rogers Ramanujan (GRR)
identities arising in the context of the AGT correspondence.

The AGT correspondence is a relation between four dimensional instanton space and
two dimensional conformal field theories \REF\AGT{L.F. Alday, D. Gaiotto, Y. Tachikawa,
Lett. Math. Phys. 91 (2010) 167.}\r\AGT. In this framework, it was proposed 
\REF\BF{V. Belavin and B. Feigin, JHEP 1107 (2011) 79}\REF\NT{T. Nishioka and Y. Tachikawa,
Phys. Rev. D84 (2011) 046009} in the papers refs. \r{\BF,\NT}\ that the coset
model
$${\cal A}(r,p)={\cal H}\times sl(p)_r\times {sl(r)_p\times sl(r)_n\over sl(r)_{n+p}},\e$$
corresponds to instanton counting on $C^2/Z_p$ where the gauge group is $U(r)$.
Here $\cal H$ stands for the Heisenberg algebra and $n$ is a free parameter
expressed in terms of Nekrasov's deformation parameters $\epsilon_{1,2}$
\REF\N{N.A. Nekrasov, Adv. Theor. Math. Phys. 7(2004) 831.}\r\N.

Our purpose here is to express characters arising in the AGT correspondence as 
GRR sums. Many such expressions for characters of conformal field theories are known
already, for example, 
\REF\Lep{J. Lepowski and M. Primc, Contemporary Mathematics vol. 46 (AMS, Providence, 1985).}
\REF\Das{V.S. Dasmahapatra, T.R. Klassen, B.M. McCoy and E. Melzer, J. Mod. Phys. B7 (1993) 3617.}
\REF\Kedem{R. Kedem, T.R. Klassen, B.M. McCoy and
E. Melzer, Phys. Lett. 307B (1993) 68.}
\REF\Baver{E. Baver and D. Gepner, Phys. Lett. B372 (1996) 231.} 
\REF\Kun{A. Kuniba, T. Nakanishi and J. Suzuki, Mod. Phys. Lett. A8 (1993) 1649.}
\r{\Lep,\Das,\Kedem,\Baver,\Kun}.

We propose that the coset ${\cal A}(r,p)$ for every positive integer $n$, and any $r$ and $p$,
affords a description of the characters as GRR sums.
We will treat here the cases of $r=2$, $n=1$ and any $p$, describing all the characters
of these models as GRR $q$ series sums.
\par
Our starting point is the coset relevant for the correspondence with 
$Z_p$, where $p=1,2,3,\ldots$, instanton moduli space, eq. (1). 
We can write the factor of $sl(p)_r$ in eq. (1) as $U(1)^{p-1}\times sl(p)_r/U(1)^{p-1}$.
We will subsequently ignore the factors of $U(1)^{p-1}$ since they enter trivially into the characters
of the theory.
Using rank--level duality, the coset eq. (1) can be written as, for $r=2$,
$$\twoline{G_{p,k_1,k_2,\ldots,k_{p},m_0,m_p}=}{\sum_{m_1,\ldots, m_{p-1}
\atop m_i+m_{i+1}=k_{i+1} \mod 2,\ i=0,1,\ldots,p-1}
\prod_{i=1}^{p-1}  {sl(2)_i^{m_{i-1}}\times sl(2)_1^{k_i}\over
sl(2)_{i+1}^{m_i}} \times {sl(2)_p^{m_{p-1}} \times sl(2)_n^{k_p}
\over sl(2)_{n+p}^{m_p}},}$$
where we denote by $sl(2)_f^s$ the affine $sl(2)$ theory at level 
$f$ and the representation with twice isospin $s$, $0\leq s \leq f$.
The indices $k_i$ and $m_0$ are $0$ or $1$ for $i=1,2,\ldots,p-1$, $k_p=0,1,\ldots,n$
and $m_p=0,1,2,\ldots,n+p$.
The equality of the coset to $G$ implies a similar relation fo the characters where each of
the factor coset is replaced by the relevant branching function. 

We will concentrate from now on the case $n=1$, $r=2$ and any non-negative integer $p$. We shall also assume for now
that $k_i=0$ for $i=1,2,\ldots,p-1$ and $m_0=0$. We shall denote $m_p=l$, $l=0,1,\ldots p+1$.
In this case the characters can be written as
$$\chi_{p,l}(q)=\sum_{m_1,m_2,\ldots m_{p-1}\ {\rm all\ even}}\prod_{i=1}^p X(m_{i-1}+1,m_{i}+1,i+2),\e$$
where $m_i$ goes from $0$ to $i$. Recall that $m_0=0$ and $m_p=l$. $k_p=0,1$ is an irrelevant
factor determined by $k_p=m_{p}\mod 2$. Here we denote by $X$ the characters of the unitary minimal
model $M_{t,t+1}$ which are given by
$$X(p,r,t)= {\Theta_{u_-(p,r),t(t+1)}(q)-\Theta_{u_+(p,r),t(t+1)}(q)\over \prod_{y=1}^\infty (1-q^y)},\e$$
where the theta function at level $h$ is defined by
$$\Theta_{w,h}(q)=\sum_{j\in {w\over 2 h}+Z} q^{h j^2},\e$$
and 
$$u_\pm(p,r)=(t+1)p\pm t r,\e$$
and $p=0,1,\ldots,t-1$ and $r=0,1,\ldots,t$.

Let us turn now to the identities. We need some notation. Define
$$S(p,\vec Q,\vec A|q)=\sum_{b_1,b_2,\ldots, b_p=0\atop \vec b=\vec Q\mod 2}^\infty {q^{\vec b C_p \vec b/4-\vec A \vec b/2}
\over (q)_{b_1} (q)_{b_2}\ldots (q)_{b_p} },\e$$
where we denoted
$$(q)_b=\prod_{c=1}^b (1-q^c),\e$$
and $C_p$ is the cartan matrix of the algebra $A_p$, $(C_p)_{a,b}=2 \delta_{a,b}-\delta_{a,b+1}-\delta_{a,b-1}$.
Here $\vec Q \mod 2$ and $\vec A$ are some integer vectors.
We define the unit vectors $e_r$ by $(e_r)_a=\delta_{r,a}$, for $r=1,2\ldots,p$ and $e_r$ is zero otherwise.
We take 
$${\vec A}_l=e_l,\e$$
and
$$\vec{Q}_l=e_1+e_3+e_5+\ldots+e_v,\e$$
where $v$ is the maximal odd integer less than $l$.
Our identities for the characters $\chi_{p,l}(q)$ are then given by
$$\chi_{p,l}(q)=q^{\Delta_{p,l}} S(p,\vec Q_l,\vec A_l|q),\e$$
where $\Delta_{p,l}$ is some dimension.
We checked this identities up to order $q^{20}$ for $p=1,2,3,4,5$ and $l=0,1,2,3,4,5$ and they are confirmed.
We conjecture that they hold for any $p=1,2,3,\ldots$ and any $l=0,1,2,\ldots,p+1$.

Now, consider the coset eq. (2) expressing $G$. All the intermediate cosets can be cancelled, since they are
summed over. We conclude that $G$ can be written as the coset,
$$G_{p,k_1,k_2,\ldots,k_{p},m_0,m_p }={sl(2)_1^{m_0}\times sl(2)_1^{k_1}\times sl(2)_1^{k_2}\ldots sl(2)_1^{k_{p-1}}
\times sl(2)_n^{k_p}\over sl(2)_{n+p}^{m_p}},\e$$
where we will specialize, as before, to the case $n=1$. 

We see from this expression for $G$, eq. (12), that the coset is invariant under any permutation of $m_0,k_1,k_2,\ldots, 
k_p=0{\ \rm or\ 1}$, and that, moreover, the following condition holds,
$$m_0+\sum_{i=1}^p k_i=m_p\mod 2.\e$$
As before, we denote my $l=m_p$. Thus, this coset depends only on the sum
$$r=m_0+\sum_{i=1}^p k_i,\e$$
and we can denote the coset $G$ as
$$G^l_r=G_{p,m_0,k_1,k_2,\ldots, k_p,m_p},\e$$
where, in addition, we have the condition
$$l=r\mod 2.\e$$
Accordingly, we denote the character of this coset as $c^l_r(q)$ and it is given by
$$c^l_r(q)=\sum_{m_1,m_2,\ldots, m_{p-1} \atop m_i+m_{i+1}=k_{i+1}\mod2,\ i=0,1,\ldots,p-1}
\prod_{i=1}^p X(m_{i-1}+1,m_i+1,i+2),\e$$
where we denote as before by $X$ the characters of the minimal models. Here we can choose any $m_0,k_1,k_2,\ldots, k_p=0,1$
as long as $r=m_0+\sum_{i=1}^p k_i$ and the result will be the same. In addition, from the field identifications
of the coset eq. (12), we find that the result is invariant under the simultaneous transformation, $m_0\rarrow 1-m_0$,
$k_i\rarrow 1-k_i$, for $i=1,2,\ldots,p$ and $m_p\rarrow p+1-m_p$. This implies that the characters obey,
$$c^l_r(q)=c^{p+1-l}_{p+1-r}(q),\e$$
for all $l$ and $r$ which are in the range, $l,r=0,1,\ldots, p+1$ 
and $l=r\mod 2$.

The invariance under permutations of $m_0,k_i$ implies a great deal of nontrivial relations for the characters of the minimal models,
and through them for the theta functions. 
Example: Assume $p=2$. 
Take $l=r=1$. Then we get,
$$c^1_1(q)=c^2_2(q)=
X(1, 2, 3) X(2, 2, 4)=
X(1, 1, 3) X(1, 2, 4) + X(1, 3, 3) X(3, 2, 4).\e$$
Another identity is for $l=3$ and $r=1$,
$$c^3_1(q)=c^0_2(q)= 
X(1, 2, 3) X(2, 4, 4)=
X(1, 1, 3) X(1, 4, 4) + X(1, 3, 3) X(3, 4, 4).\e$$
These are all the nontrivial identities for $p=2$.
For higher $p$ we have a great deal more identities, which we do not list for brevity.
The theta function identities obtained this way, which we proved by the 
coset construction, are  
highly nontrivial and there does not seem an obvious elementary 
way to prove them.

We are now ready to give the GRR expression for the characters
$c^l_r(q)$. Define, as before
$$\vec Q_l=e_1+e_3+\ldots+e_v,\e$$
where $v$ is the largest odd integer less than $l$. Also define
$$\vec Q_{l,r}=Q_l+e_{r-1}+e_{r-3}+\ldots.\e$$
Due to the field identification $c^l_r(q)=c^{p+1-l}_{p+1-r}(q)$
we can assume, without loss of generality, that $l>r$ or
$l=r\leq (p+1)/2$. Then we have the following expression for the
characters,
$$c^l_r(q)=q^{\Delta_{l,r}} S(p,\vec Q_{l,r},e_l|q),\e$$
where $\Delta_{l,r}$ is some dimension and $S$ was defined by eq. (7).
We checked this identities for $p=1,2,3,4,5$ and all $l$ and $r$
up to order $q^{20}$ using a Mathematica program and they are, indeed,
obeyed.
 
In fact, using rank level duality, the case of $n=1$ of the coset, eq. (2) can be written as
$$sl(p+1)_2\over U(1)^p,\e$$
which is a theory of generalized parafermions. A conjecture for the GRR expression for the characters of the pure parafermions sector
of this model was made in ref. \r\Kun. Our results are in agreement with this conjecture
and extend it to all the fields in the theory.

Example: Take $p=2$, $l=r=0$. We then obtain the identity, 
according to eq. (23)
$$c^0_0(q)=q^{1/30}\sum_{b_1,b_2=0\atop b_1,b_2={\rm even}}^\infty
{q^{(b_1^2+b_2^2-b_1 b_2)/2} \over (q)_{b_1} (q)_{b_2}}.\e$$
Take $p=2$ and $l,r=1$ then we get the identity from eq. (23)
$$c^1_1(q)=q^{2/15} \sum_{b_1,b_2=0\atop b_1-1=b_2=0\mod 2}^\infty
{q^{(b_1^2+b_2^2-b_1 b_2-b_1)/2}\over (q)_{b_1} (q)_{b_2}}.\e$$

At this point, the reader might wonder what happens to $S(p,\vec Q,e_l|q)$
when $\vec Q$ is not equal to one of the $\vec Q_{l,r}$, eq. (23). The answer is
quite surprising:
 
{\bf Conjecture:} 
For any choice of $\vec Q$ the sum $S(p,\vec Q,e_l|q)$ is equal
to one of the characters $c^l_{r_Q}(q)$ for some $r_Q=0,1,2,\ldots,p+1$ and
$r_Q=l\mod 2$.

We checked this conjecture, by direct calculation, 
for $p=1,2,3,4,5$ and all $l$ and $\vec Q$ up to order $q^{20}$ and 
it works.

In the case of $p=1$ we recover the known identities, 
\REF\Slater{L.J. Slater, Proc. London Math. Soc. 54 (1953) 147.}
refs. \r{\Slater,\Kedem}, 
for the Ising model,
$$c^0_0(q)=c^2_2(q)=\sum_{b=0\atop b\ {\rm even}}^\infty {q^{b^2/2}\over (q)_b},\e$$
$$c^0_2(q)=c^2_0(q)=\sum_{b=0\atop b\ {\rm odd}}^\infty {q^{b^2/2}\over (q)_b},\e$$
$$c^1_1(q)=\sum_{b=0\atop b {\rm\ even}}^\infty {q^{(b^2-b)/2}\over (q)_b}=
\sum_{b=0\atop b {\rm\ odd}}^\infty {q^{(b^2-b)/2}\over (q)_b},\e$$
where the last identity is due to the conjecture. 
We ignored factors of $q^\Delta$. These are indeed all the possibilities, in this case,
for $l$ and $\vec Q$.

In the case of $p=2$ we find from eq. (23),
$$c^0_0 (q) q^{-1/30}=S (2,0,0|q),\e$$
$$c^0_2(q) q^{-8/15 + 1/2}=S(2,e_1,0|q)= S(2,e_1+e_2,0| q),\e$$
$$c^1_1(q) q^{-2/15}=S (2,0,e_1|q)=S(2,e_1,e_1|q)=S(2,e_1+e_2,e_1|q),\e$$
$$c^1_3 (q) q^{-19/30 + 1/2}=S (2,e_2,e_1|q).\e$$
These are all the possibilities for $l$ and $\vec Q$, agreeing with the conjecture. We see that our identities
generalize Slater's identities for the Ising model.

For $p=3$, we find the following identities,
$$c^0_0(q) q^{-1/24}=S (3,0,0|q),\e$$
$$\twoline{c^0_2(q) q^{-13/24 + 1/2}= 
S(3,e_ 1,0|q)=}{ 
  S(3,e_ 1+e_2,0|q)=S(3,e_1 + e_2 + e_3,0|q)=S(3,e_2,0|q),}$$
$$c^0_4(q) q^{-25/24 + 1}=S(3,e_ 1+e_3,0|q),\e$$
$$\twoline{c_1^1(q) q^{-1/6}= 
 S(3,0,e_ 1|q)= 
  S(3,e_1,e_1|q)=}{S(3,e_1+e_2,e_ 1|q)=
    S(3,e_1+e_2+e_3,e_1|q),}$$
$$\twoline{c^1_3(q) q^{-2/3 + 1/2} = 
 S (3,e_2,e_1|q)= 
  S(3,e_3,e_1|q) =}{ 
   S(3,e_1+e_3,e_1|q)=S(3,e_ 2+e_3,e_1|q),}$$
$$c^2_0(q) q^{-17/24 + 1/2}=S(3,e_1,e_2|q),\e$$
$$\twoline{c^2_2(q) q^{-5/24} = 
 S(3,0,e_2|q)= 
  S(3,e_2,e_2|q) =}{ 
   S (3,e_1+e_2, e_2|q)= 
    S(3,e_1 + e_3, e_2 | q) = S (3, e_1 + e_2 + e_3, e_2 | q).}$$
Again, this agrees with the conjecture.

We checked also the cases of $p=4,5$ and, indeed, the conjecture holds. The results are to long to list here.

Recall that we defined the expression for $S(p,\vec Q,\vec A|q)$, eq. (7), by using the Cartan matrix
of the algebra $A_p$. We may generalize this expression using the Cartan matrix of any of the
simply laced algebras, which are of the types, $G=A_p,D_p,E_6,E_7,E_8$. Thus define
$$S(G,\vec Q,\vec A|q)=\sum_{b_1,b_2,\ldots,b_p\atop \vec b=\vec Q\mod2} {q^{\vec b C_G\vec b/4-\vec A\vec b/2}
\over (q)_{b_1} (q)_{b_2}\ldots (q)_{b_p}},\e$$
where $C_G$ is the Cartan matrix of the algebra $G$ and we take $\vec A=e_l$ for some $l=0,1,\ldots,p$.
As before, $\vec Q$ is taken to be any vector of length $p$ of integers modulo $2$.

We observe immediately that for any $G$, 
$S(G,\vec Q,\vec A|q)$ has the correct form to be characters of some conformal
field theory. Namely, for any $l$ and $\vec Q$ it is of the form,
$$S(G,\vec Q,e_l|q)=q^\delta \sum_{i=0}^\infty a_i q^i,\e$$
where $a_i$ are some nonnegative integers. 

For example, take $G=D_4$. The Cartan matrix is
$$C_{D_4}=\pmatrix{2&1&1&1\cr 
                   1&2&0&0\cr
                   1&0&2&0\cr
                   1&0&0&2\cr}.\e$$
For $\vec Q=0$ and $l=0$ we find
$$S(D_4,0,0|q)=
1+12 q^2+27 q^3+87 q^4+189 q^5+463 q^6+954 q^7+2013 q^8+\ldots.\e$$
This looks exactly like the character of the identity field in some conformal field theory.
Similarly, other $S(D_4,\vec Q,e_l|q)$ have the form of the characters of other fields.
We thus believe that for any of the simply laced Lie algebras, $G$, these sums express the
characters of some conformal field theory, just as in the case of $A_p$ discussed previously.
Actually, the conjecture of ref. \r\Kun\ implies that at least for $l=0$ these are the characters of the coset,
$${G_2\over U(1)^r},\e$$
where $r$ is the rank and $G_2$ is the WZW model with the algebra $G$ at level $2$.
We believe that the rest of the characters, $l>0$, are given by $S(G,\vec Q, e_l|q)$, eq. (41), for some $l$ and $\vec Q$.

We can check the equivalent of our conjecture for the other algebras. This implies
that some $l$ for each $\vec Q$ we should find a small number of results, giving identities
for the sum $S(G,\vec Q,e_l|q)$ among themselves, which are highly nontrivial. We call these identities
generalized Slater (GS) identities.

To exemplify, take again $G=D_4$. For $l=0$ we find,
$$\twoline{S (G, e_ 1, 0 | q) = 
 S (G, e_ 1 + e_ 2, 0 | q) = 
  S (G, e_ 1 + e_ 2 + e_ 3, 0 | q) =}{ 
   S (G, e_ 2 + e_ 3 + e_ 4, 0 | q) = 
    S (e_ 1 + e_ 2 + e_ 3 + e_ 4, 0 | q),}$$
and all together there are three different characters with $l=0$.

For $l=1$ we find
$$\twoline{S (G, 0, e_ 1 | q) = 
 S (G, e_ 1, e_ 1 | q) = 
  S (G, e_ 1 + e_ 2, e_ 1 | q) =}{ S (G, e_ 2 + e_ 3, e_ 1 | q) =
    S (G, e_ 1 + e_ 2 + e_ 3, e_ 1 | q) = 
     S (G, e_ 1 + e_ 2 + e_ 3 + e_ 4, e_ 1 | q),}$$
and altogether there are again three different characters with $l=1$.

We checked these identities for other groups and other $l$ and indeed we get a great deal of
nontrivial GS identities.

\ack
We thank A. Babichenko and B. Feigin for useful discussions and Ida Deichaite for remarks.

\refout
\bye

\bye